PERSPECTIVE

# Underwater imaging without colour distortions requires RAW capture

Derya Akkaynak[1,2] | Michael S. Brown[3]

[1]Hatter Department of Marine Technologies, Leon H. Charney School of Marine Sciences, University of Haifa, Haifa, Israel
[2]Interuniversity Institute of Marine Sciences in Eilat, Eilat, Israel
[3]Department of Electrical Engineering and Computer Science, York University, Toronto, Ontario, Canada

**Correspondence**
Derya Akkaynak
Email: dakkaynak@univ.haifa.ac.il

**Funding information**
HORIZON EUROPE Framework Programme, Grant/Award Number: 101094924; Office of Naval Research Global, Grant/Award Number: N629092312104; Schmidt Marine Technology Partners, Grant/Award Number: G-22-63208 and G-24-66610; Israel Science Foundation, Grant/Award Number: 1055/22 and 2788/22

**Handling Editor:** Aaron Ellison

## Abstract

1. Consumer cameras are not designed to be scientific instruments, yet they are increasingly used in science not just to see, but to measure. This is possible because camera sensors respond approximately linearly to the light they receive, a property that can, in principle, support quantitative measurements of colour.
2. Whether that quantitative measurement potential is preserved depends on what happens to the image after capture. Minimally processed sensor images, commonly known as RAW, retain the approximately linear response needed. Instead of RAW, however, the default output of most cameras is a heavily processed image optimized for visual appearance, compact storage and immediate use, typically saved as a JPEG file.
3. The convenience offered by an in-camera processed image comes at a scientific cost. RAW capture requires more storage and processing effort but leaves calibration choices open. Processed images reduce those demands at acquisition and in post-processing but transfer irreversible decisions about colour manipulation from the researcher to the camera.
4. The preservation of RAW sensor data is necessary but not sufficient for quantitative colour measurement. Pixel values must also be calibrated to account for the light conditions under which the image was acquired. Underwater, this calibration is more challenging than in clear air, because light is absorbed and scattered as it travels through water, resulting in effects that depend on both wavelength and distance. Consequently, colour distortions vary across the scene and generally cannot be corrected with a single global colour transformation like white balancing; instead, calibration requires information about the imaging distance. We explain the consequences for quantitative imaging and introduce the 3P Protocol for distortion-free colour imaging underwater: **Preserve** RAW sensor data, **place** a known colour reference in the scene and **plan** acquisition to control or measure imaging distance.
5. Calibration and reconstruction methods will evolve but cannot recover information discarded at acquisition. Preserving the measurement today keeps image datasets open to the questions and methods of tomorrow.

## 1 | INTRODUCTION

Digital images and video support a wide range of ecological studies, including increasingly diverse underwater applications. They facilitate assessments of presence–absence, abundance, diversity and condition (King et al., 2020; Rovero & Kays, 2021; Schramm et al., 2020), enable exploration of extreme environments (Bell et al., 2022; Griffiths et al., 2021), help create 3D maps (Monfort et al., 2021; Pedersen et al., 2025) and document rarely observed animal behaviours (Allen et al., 2017; Maalouf et al., 2026; Shlesinger et al., 2021). In many of these applications, images are used as visual records for which the presence, identity and relative differences among features are sufficient for interpretation. They can also serve as measurements, for example, when quantifying coral bleaching intensity (Bahr et al., 2020; Siebeck et al., 2006), estimating pigment concentration (Ferrara et al., 2026; Winters et al., 2009) or assessing animal coloration (Arenas et al., 2014; Marom et al., 2025). In such cases, cameras must be treated as measurement devices and images must be acquired in a manner that preserves the information needed for calibration and quantitative analysis, including analyses that may not have been anticipated at the time of acquisition.

This need for calibration arises because a camera image is not a direct or objective record of scene colour. The pixel intensities recorded by a camera are measurements made by that sensor under a particular illumination; they are not intrinsic properties of the photographed scene. This is because cameras sense spectral radiance (incoming light from a scene towards the camera) through sensor filters with wavelength-dependent sensitivities across the visible spectrum. In most cameras, three filter classes provide red-, green- and blue-weighted (RGB) responses, and their exact specifications vary by manufacturer. The recorded RGB values are device- and illumination-dependent (Figure 1a). Cameras can preserve these measurements with relatively little processing in their native RAW image formats (e.g. Sony's. ARW, Nikon's. NEF). In this form, the pixel values remain approximately linear with sensor exposure and scene radiance over their usable, unsaturated range (Kim et al., 2012). RAW images therefore retain the information needed for calibration, allowing quantitative colour measurements to be derived from their pixel values.

In addition to RAW images, consumer cameras also generate—and often output by default—images that have undergone extensive in-camera processing (Figure 1b). These pipelines include operations such as demosaicing, white balancing, tone mapping, noise reduction, sharpening and colour rendering, which are designed to produce visually pleasing images while balancing image quality, processing speed and file size (Brown, 2023). The resulting files, typically still images in Joint Photographic Experts Group (JPEG) format and videos in Moving Picture Experts Group 4 (MP4) format, are encoded in a display-ready RGB colour space such as *standard RGB* (sRGB). Unlike RAW data, however, the pixel values in these processed images no longer have a predictable relationship to the original sensor measurements. Importantly, because these proprietary, undocumented operations cannot be undone, the original RAW data cannot be recovered from in-camera processed images.

RAW images are therefore a necessary starting point for quantitative colour analysis, but they do not themselves provide calibrated colour. Rather, they preserve the information needed for calibration, which has two goals: (1) standardizing illumination and (2) transforming the resulting device-dependent RGB values into a device-independent representation. Only then can colour values acquired across different cameras, times, locations and light conditions be meaningfully compared. For images taken in clear air, well-established protocols describe these steps (Akkaynak et al., 2014; Johnsen, 2016; Stevens et al., 2007; Troscianko & Stevens, 2015). Because clear air is non-attenuating, a global correction—typically white balancing using a colour chart—is generally sufficient to standardize illumination, provided the scene is reasonably uniformly lit. Transformation to a device-independent colour space is done following established colorimetry standards by first mapping the camera's RGB values to the Commission Internationale de l'Éclairage (CIE) XYZ colour space under a specified illuminant and then converting them to a standard RGB space (Schanda, 2007).

Unlike clear air, water is an attenuating medium. Light is strongly absorbed and scattered as it propagates through water, making calibration dependent not only on the camera, but also on the optical properties of water, the depth of the camera below the surface and the depth of the scene (i.e., the imaging distance) (Akkaynak et al., 2017; Akkaynak & Treibitz, 2018). Quantitative colour recovery therefore requires methods that account for distance- and wavelength-dependent effects (Figure 1c). A colour chart alone is generally insufficient to correct for these effects (Akkaynak & Treibitz, 2019).

In this Perspective, we explain why underwater colour cannot be standardized using the global corrections that work in clear air, and why quantitative colour capture instead requires capturing and preserving raw sensor measurements, placing a known colour reference in the scene and planning image acquisition to measure or control the camera-to-scene distance, which we summarize as the 3P Protocol. The stakes are high: data not captured as RAW is irretrievably lost; no future advance in calibration or reconstruction can restore what in-camera processing has already destroyed. We therefore outline steps for researchers and manufacturers to acquire and archive underwater imagery as reusable quantitative data, rather than locking observations into the irreversible processing applied by cameras at acquisition.

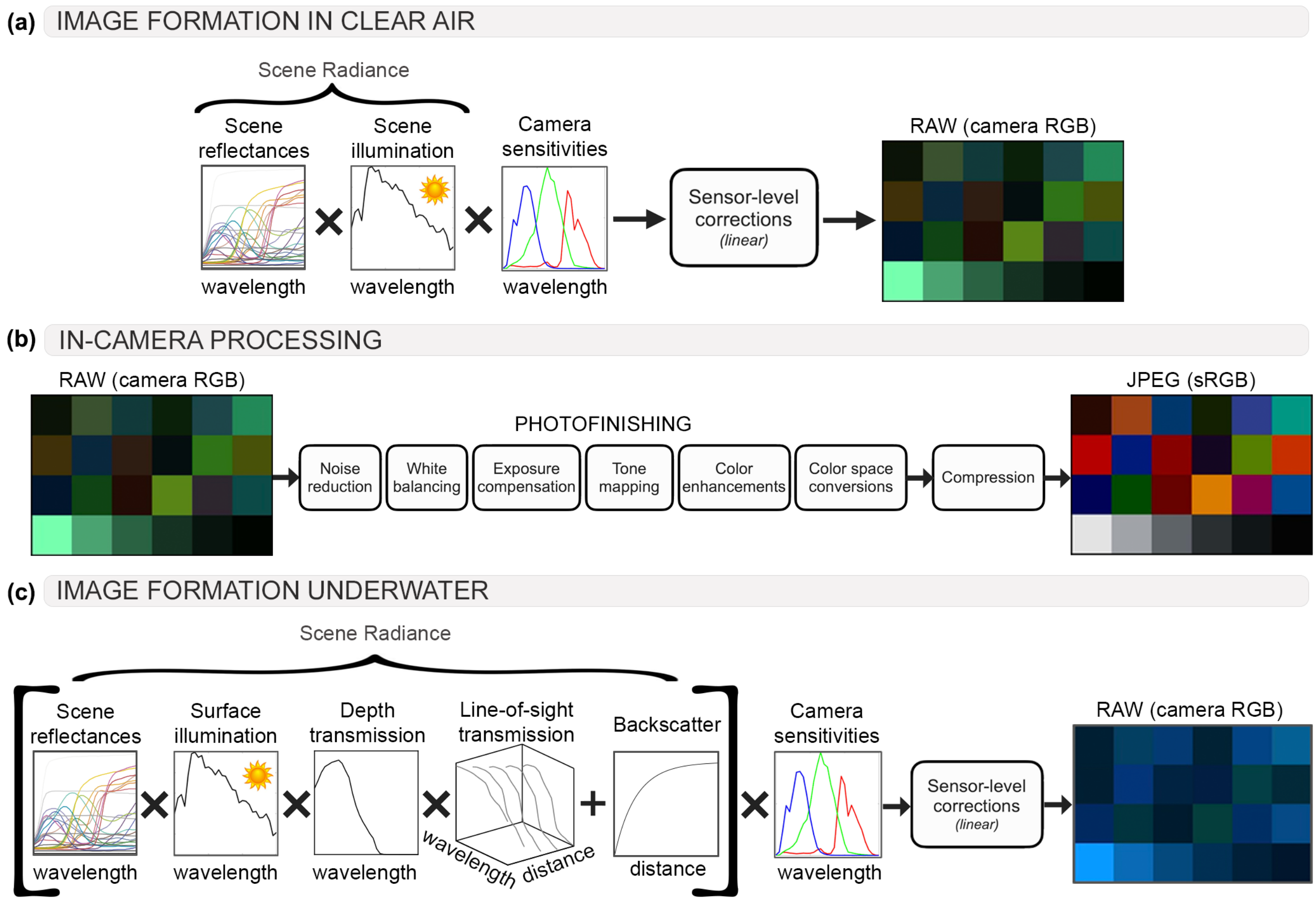

FIGURE 1 From scene radiance to camera pixel values in air and underwater. (a) A camera records scene radiance through sensor-specific spectral sensitivities, producing RAW RGB values that retain an approximately linear relationship with sensor exposure over the usable range. (b) In-camera processing transforms these measurements to produce a small-sized display-ready image. Because these operations and their parameters vary among cameras, the resulting pixel values no longer have a predictable relationship to original sensor measurements. (c) Underwater, image formation additionally depends on wavelength- and distance-dependent attenuation and backscatter. Consequently, recovering comparable scene colours requires accounting for both water properties and camera-to-scene distance.

## 2 | SNAPSHOT OF THE PROBLEM FOR SCIENTISTS

For still photography, quantitative colour measurement generally requires capturing RAW images, which carries practical costs (Table 1). Video presents the same underlying requirement, although access to minimally processed sensor data for video is less common and workflows differ from still photography. RAW files are larger in terms of storage and take more time to save, transfer and process. RAW image pixel values cannot be directly interpreted or compared between cameras because they represent sensor-specific measurements. Converting them to a standard colour space requires some understanding of digital image formation and access to appropriate tools and expertise. RAW processing also increases computational and data handling demands, particularly for large image datasets, and camera-specific RAW formats or processing workflows can complicate long-term data management and reproducibility. RAW capture must also be deliberately incorporated into the imaging workflow rather than assumed to occur automatically. Thus, capturing RAW increases storage and processing demands, and adds workflow complexity, but preserves the information needed for calibration and leaves calibration and analysis choices open for later.

In-camera processed imagery offers an attractive trade-off: smaller, display-ready files and broad compatibility. This convenience, however, comes at the cost of control over how the recorded sensor values are transformed. Tone mapping and gamma encoding, for example, alter the relationship between pixel values and scene radiance; in-camera colour rendering deliberately shifts hue and saturation; white balancing changes the relative magnitude of the colour channels; and noise reduction and sharpening modify spatial information. Processed images are also typically compressed, which can introduce artefacts even at 'high-quality' settings (Box 1). Thus, while in-camera processing saves resources at the time of acquisition and in post-processing, it transfers decisions about how colours are transformed from the researcher to the camera, where they are often irreversible. Therefore, the challenge is not simply that photographs look different. Because the operations, their parameters and their ordering vary among manufacturers and camera models, JPEG (Joint Photographic Experts Group) refers both to a widely used image file format and to the lossy compression algorithm used to reduce file size. JPEG compression works by permanently discarding image information that is expected to be minimally perceptible to the human visual system, allowing image files to be an order of magnitude smaller than the original with little apparent loss in visual quality.

**TABLE 1** Scientific and practical trade-offs between RAW and in-camera processed imagery.

| Property | RAW | In-camera processed image/video |
|---|---|---|
| Relationship to sensor measurement | Pixel values retain an approximately linear relationship with sensor exposure over the usable, unsaturated range | Relationship to the original sensor measurement is altered by the image processing pipeline |
| Colour representation | Device-dependent; requires calibration before comparison across cameras | Typically encoded in a standard display colour space, but values have already been modified by camera-specific processing |
| Quantitative colour analysis | Suitable after appropriate calibration | Generally unsuitable when absolute or comparable colour measurements are required |
| Future reprocessing | Preserves information needed for alternative or improved calibration or processing | Original sensor measurements generally cannot be uniquely recovered |
| Cross-camera comparability | Can be achieved through calibration | Not guaranteed despite use of a common output colour space. Cameras apply proprietary colour manipulation to impart a particular 'look and feel' |
| Storage and transfer | Larger files; greater storage and transfer requirements and associated costs | Smaller files; easier storage, transfer, and sharing |
| Processing requirements | Requires manipulation/calibration before quantitative use or conventional display | Immediately viewable and widely compatible |
| Appearance | Unprocessed rendering may initially appear dark and unfamiliar | Optimized for visually pleasing colour and contrast |
| Format/metadata standardization | Camera-dependent formats and metadata can complicate long-term workflows | Widely standardized formats and broad software support |
| Appropriate use | Quantitative colour measurements, datasets whose future colour use cannot be ruled out | Presence–absence, identification, classification, documentation and other applications where pixel values are not themselves measurements |

**BOX 1 JPEG files versus JPEG compression: which is the problem?**

JPEG compression itself is not the primary obstacle to quantitative colour analysis. If a linear RAW image were compressed directly as a high-quality JPEG without any other modifications, its pixel values would remain proportional to scene radiance (apart from small quantization and compression artefacts) and could still, in principle, be used for quantitative colour measurements.

The scientific limitation arises because consumer cameras apply a sequence of photofinishing operations *before* JPEG compression (Figure 1b). These operations alter the linear relationship between pixel values and scene radiance and are generally undocumented and cannot be reversed. Consequently, even a high-quality JPEG that has undergone in-camera processing cannot be uniquely converted back to the original RAW image or used as a reliable quantitative colour measurement.

processed pixel values no longer provide a known or reproducible representation of the original sensor measurements. Simply stated, JPEG images or in-camera processed imagery in other formats cannot be used as reliable measurements.

## 3 | RAW DATA CANNOT BE RECOVERED FROM IN-CAMERA PROCESSED IMAGES

A common misconception is that in-camera processing can be reversed. For example, because gamma encoding, which is used in converting a linear RGB image to sRGB, is a well-defined mathematical transformation, it is sometimes assumed that applying the inverse gamma function restores the linear relationship between pixel values and scene radiance or even reconstructs the original RAW image. However, gamma encoding is only one step within a much larger processing pipeline. Simply undoing the gamma function cannot recover the original image because the other operations in the pipeline, their parameters and often their ordering are unknown. The presence of a colour chart in the image does not solve this problem. A colour chart is a reference, not a correction model; it can be used to adjust the appearance of an in-camera processed image against known colours, but it cannot reveal the unknown sequence of operations that produced the image or invert the camera pipeline, both of which are needed to reconstruct the original sensor measurements.

Computational methods, including computer vision and machine learning approaches, can estimate plausible RAW images under favourable, typically clear air conditions (Afifi & Abuolaim, 2021; Afifi et al., 2022; Conde et al., 2023). However, these methods learn statistical approximations rather than recover the original sensor data, and this limitation is not simply a matter of training data or model sophistication: because the mapping from RAW to processed image is fundamentally one-to-many, no model, however well trained, and however underwater-specific, can guarantee recovery of the original values.

## 4 | MAKING UNDERWATER COLOURS COMPARABLE

Consider three common scenarios in aquatic sciences: (1) photographing the same subject (e.g. a coral colony) over time with the same camera, (2) photographing the same subjects at different sites over time with the same camera, and (3) photographing the same species across regions and over time with different cameras. How can comparability be ensured in each case? Although their logistical complexities differ, the underlying measurement problem is the same for all scenarios: separating changes in the subject from changes introduced by variations in illumination, water properties, imaging geometry, and the imaging system itself. In practice, a study designed around a single camera may ultimately rely on multiple cameras, as equipment can fail, flood or need to be replaced during long-term data collection.

The calibration goals are the same underwater as in clear air: (1) standardize illumination and (2) transform device-dependent RGB values into a standard colour space. However, underwater, standardizing illumination first requires accounting for how water alters light as it propagates through the medium (Figure 1c). This requires careful planning prior to image acquisition and appropriate data collection during image acquisition.

Accounting for the water medium. The most fundamental way in which underwater image formation differs from that in clear air is the interaction of light with the medium (Akkaynak & Treibitz, 2018; Mobley, 1994). Two processes are particularly important: attenuation, the wavelength-dependent loss of light caused by absorption and scattering, and backscatter, light scattered by particles in the water column into the camera from directions other than the subject. Light is attenuated both as it travels from the source to the scene and as it travels from the scene to the camera (Figure 1c), but attenuation along these two paths has different effects on the recorded image. Under natural illumination, attenuation with water depth gradually narrows the spectrum of the transmitted light reaching the scene, often producing a dominant blue or green hue, although other colours can dominate depending on the water constituents. Along the line-of-sight, attenuation further weakens and spectrally alters the light transmitted from the scene, while backscatter adds a layer of coloured haze that reduces contrast and obscures details.

Because the camera-to-scene distance varies across an image, attenuation and backscatter also vary from pixel to pixel. Underwater illumination standardization is therefore inherently distance-dependent (Akkaynak & Treibitz, 2018). When the camera is close to the scene, attenuation and backscatter along the line-of-sight are minimized, maximizing the recoverable scene signal (Figure 2a). As the imaging distance increases, light travels through more water before reaching the camera: colours fade, backscatter intensifies and the amount of useful, recoverable signal decreases rapidly. At sufficiently large distances, the image contains little recoverable information about scene colour and becomes dominated by backscatter; a practical rule of thumb in underwater imaging is, therefore, to get as close to the subject as possible.

Distance-dependent effects cannot generally be corrected using global methods such as white balancing. Correcting these effects, therefore, requires knowing how far the camera is from each point in the scene. This camera-to-scene distance can be represented by a pixelwise distance map,[1] which can be obtained, for example, from multiple overlapping images using single-camera methods such as structure-from-motion (Akkaynak & Treibitz, 2019), multiple-camera setups (Berman et al., 2021) or specialized ranging sensors such as lidar (Akkaynak et al., 2022).

Given a depth map and an image formation model (see, e.g. Akkaynak & Treibitz, 2018), attenuation and backscatter can be estimated and corrected for. The details of these procedures are beyond the scope of this Perspective; we refer readers to the computer vision literature for physics-based methods (Akkaynak & Treibitz, 2019; Levy et al., 2023; Mualem et al., 2026). A colour chart alone cannot correct for these distance-dependent effects. Once they have been accounted for, however, the chart provides the reference needed to standardize the recovered colours for the global illumination at that water depth.

Colour transformation. Once the effects of water have been accounted for and illumination standardized, the same colorimetric transformation used in clear air can be applied: device-dependent RGB values are converted to a device-independent colour space so that measurements are interpretable and reproducible across cameras. The colour transformation itself is therefore not fundamentally different underwater; what differs is that the RGB values must first be corrected for the effects of light propagation.

## 5 | WHAT IS THE WAY FORWARD?

Not every scientific application of underwater imagery requires the same degree of calibration. For presence–absence,

[1]A distance map is also referred to as a range map, depth map or scene depth; here, we use 'depth' to mean distance from the camera pixel to each point in the scene, not water column depth.

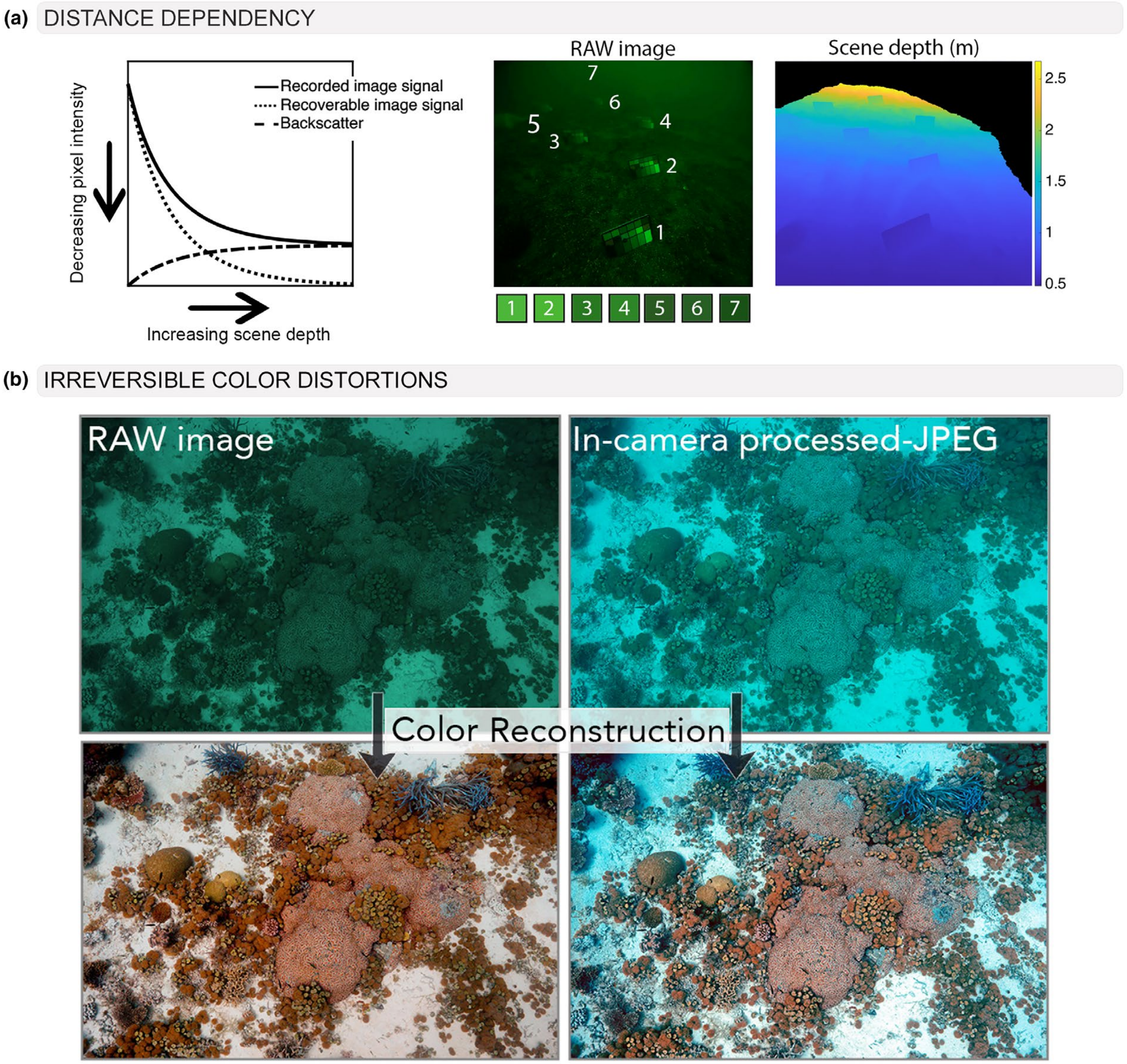

FIGURE 2 Increasing camera-to-scene distance progressively reduces recoverable scene information underwater. (a) With increasing distance, wavelength-dependent attenuation weakens the recoverable scene signal while backscatter contributes an increasing fraction of the recorded radiance. Because these effects depend on camera-to-scene distance, a single global correction such as white balancing cannot generally standardize illumination across an underwater scene. This distance dependency is illustrated by the RGB values extracted from the white patches of six colour charts positioned at different distances. A seventh point shows the saturated colour of backscatter, towards which all scene colours converge at large distances. Physics-based recovery therefore requires information about scene distance whenever line-of-sight attenuation and backscatter cannot be neglected. (b) Quantitative colour reconstruction, achieved here using the Sea-thru algorithm (Akkaynak & Treibitz, 2019), requires minimally processed sensor measurements because in-camera processing irreversibly alters the relationship between recorded pixel values and the measurements needed for subsequent physics-based correction.

identification or many classification tasks, in-camera processed imagery may be sufficient because pixel values are not themselves the measurements of interest. Therefore, the imaging workflow should be determined by the research question and the degree of quantitative comparability required. However, when quantitative colour is, or may later become, relevant, capturing and archiving minimally processed sensor data in the form of RAW files preserves the greatest flexibility for calibration, analysis and reinterpretation as methods evolve. This is particularly important for long-term ecological monitoring and datasets that may ultimately combine imagery acquired with different cameras, at different times or under different environmental conditions.

The practical requirements for preserving quantitative colour information underwater are summarized by the 3P Protocol and checklist (Box 2): **Preserve** the RAW sensor data, **place** a known colour reference in the scene, and **plan** to acquire images in a way that allows scene distance to be controlled or estimated. Together, these provide the information needed to account for the camera, illumination and distance-dependent effects of light propagation in water.

### BOX 2 The 3P Protocol and checklist for distortion-free colour imaging underwater

1. **Preserve** the RAW sensor data.
2. **Place** a colour reference in the scene.
3. **Plan** to control or measure camera-to-scene distance.

**Checklist:**

Choose suitable equipment (repeat for each camera)

- ☐ Verify that the camera can save RAW images.
- ☐ Verify that the sensor response is linear.
- ☐ Ensure that the camera's spectral sensitivities are available; if not, measure them.

Prepare and maintain colour charts

- ☐ Label or number each chart.
- ☐ Measure reflectance spectra of each chart periodically.
- ☐ Rinse charts after use and store them away from direct sunlight.

Plan your field survey

- ☐ Configure the camera to record RAW images.
- ☐ Include one or more calibrated colour charts in the scene, facing the camera and unobstructed; note chart label/number.
- ☐ Keep imaging distances as short and uniform as possible.
- ☐ For scenes imaged from a distance, plan data collection so that a pixelwise depth map can be obtained.

Handle RAW data correctly

- ☐ Convert RAW images to a linear format (e.g. TIFF) using software that does not apply additional enhancements.

Remove the effects of water

- ☐ Using the depth maps and the colour chart(s), follow physics-based methods to remove backscatter and reverse attenuation.

Transform colours to a standard colour space

- ☐ Following standard colorimetry workflows, map camera RGB values to a device-independent RGB space.

Archive files for future reuse and reproducibility

- ☐ Save all RAW images.
- ☐ Save metadata and calibration data, including spectral sensitivities and linearity calibration.
- ☐ Document all processing steps (software, versions, settings), including any custom code or toolboxes and camera firmware version.
- ☐ When available, also document field survey metadata (e.g. water depth, sky conditions, visibility, light spectrum, attenuation coefficients).
- ☐ Save final images resulting from your processing.

As with any scientific instrument, cameras used for quantitative measurement require appropriate characterization, while images require appropriate acquisition, calibration and archival. For still images, researchers should use cameras that provide access to RAW sensor data and verify that the sensor response is sufficiently linear for the intended analysis. Native RAW files should be retained together with the metadata and calibration information required to interpret them. They can then be processed directly or converted to standardized formats, provided that the conversion preserves the original linear sensor data. ISO 12234-4 (International Organization for Standardization, 2026), which defines Digital Negative (DNG) as a standardized format for RAW imagery, provides a framework for improving the interoperability and long-term accessibility of RAW data. Importantly, standardizing a file format does not standardize the camera: It neither removes camera-specific sensor characteristics and preprocessing nor converts device-dependent measurements into calibrated, device-independent colour values. Therefore, DNG can facilitate preservation and processing, but it does not replace camera-specific calibration or a documented processing workflow.

Scene distance presents an additional acquisition challenge underwater. Physics-based correction requires information about the distance between the camera and the scene at each pixel, which may require multiple overlapping images from structure-from-motion or dedicated ranging hardware. However, a simple and often overlooked strategy is to minimize the problem at acquisition: photograph the scene from short and relatively uniform distances whenever the scientific question permits (Figure 2a). How short is short?

There is no universal distance threshold because the relevant scale depends on the optical properties of the water. At camera-to-scene distances that are short relative to the attenuation properties of the water body, line-of-sight attenuation and backscatter are reduced, and global illumination corrections may become reasonable approximations. Even in this simpler case, images must still be acquired RAW with a colour chart in the scene.

When the scene distance cannot be neglected, it must instead be measured or estimated before physics-based colour recovery can be performed. This requirement exposes an important limitation of conventional in-camera processing: colour and contrast enhancements are applied before the distance-dependent effects of water are accounted for, irreversibly altering information required for subsequent reconstruction (Figure 2b). Image enhancement methods or physical interventions such as coloured lens filters, which alter appearance without accounting for scene distance, should therefore be interpreted according to what they provide: They may improve visual appearance, but they do not constitute physics-based colour standardization.

The broader technological opportunity is therefore not simply better underwater image enhancement but imaging pipelines that preserve quantitative information while retaining the practical advantages of consumer cameras. This is particularly relevant for video, where high frame rates, data rates and storage constraints often necessitate immediate processing and compression. Importantly, photofinishing and compression need not be coupled. A minimally processed 'linear-video' mode could preserve the relationship between the pixel values and the radiance of the scene while still allowing compression, enabling subsequent physics-based calibration. The same principle could apply to still images through compressed, minimally processed 'linear-JPEG' output. Such images do not have to be visually pleasing at acquisition: display rendering can be applied later, whereas discarded sensor information cannot be recovered. Compression may introduce other artefacts (e.g. quantization, chroma subsampling), but these limitations are fundamentally different from undocumented nonlinear transformations that destroy the relationship between pixel values and the original measurements.

The path forward is therefore both methodological and technological. Researchers can preserve the information required for quantitative analysis by treating cameras as measurement devices and planning image acquisition accordingly. Camera manufacturers can make this substantially easier by providing access to minimally processed sensor measurements in practical still-image and video formats. The enduring principle is simple: preserve the measurement at acquisition and postpone the irreversible rendering until later. Calibration and reconstruction methods will evolve, but they can only operate on the information that was retained in the first place.

## AUTHOR CONTRIBUTIONS

Authors conceived the ideas together and shared the writing of the manuscript. Both authors contributed critically to the drafts and gave final approval for publication.

## ACKNOWLEDGEMENTS

We thank Dr. Tom Shlesinger, COLOR (Computational Imaging and Light in the Ocean Realm) Lab members, students and teaching assistants of the Underwater Colorimetry course for helpful discussions, Mai Bonomo and Drs. Herdís GR Steinsdóttir and Grigory Solomatov for providing insightful comments, Yasemin Shlesinger for support and Marianne Fox for copyediting. This work was supported by grants to D.A. from the Schmidt Marine Technology Partners (G-22-63208, G-24-66610), Israel Science Foundation (1055/22, 2788/22), Office of Naval Research Global (N629092312104) and European Union's Horizon 2020 research and innovation program GA (101094924, ANERIS).

## CONFLICT OF INTEREST STATEMENT

Authors declare no conflict of interest.

## PEER REVIEW

The peer review history for this article is available at https://www.webofscience.com/api/gateway/wos/peer-review/10.1111/2041-210x.70417.

## DATA AVAILABILITY STATEMENT

There is no code or data associated with this Perspective article.

## ORCID

*Derya Akkaynak* https://orcid.org/0000-0002-6269-4795
*Michael S. Brown* https://orcid.org/0000-0002-9840-0795

## REFERENCES

Afifi, M., Abdelhamed, A., Abuolaim, A., Punnappurath, A., & Brown, M. S. (2022). CIE XYZ net: Unprocessing images for low-level computer vision tasks. *IEEE Transactions on Pattern Analysis and Machine Intelligence*, *44*(9), 4688–4700. https://doi.org/10.1109/TPAMI.2021.3070580

Afifi, M., & Abuolaim, A. (2021). Semi-supervised raw-to-raw mapping. Proceedings of the British Machine Vision Conference (BMVC). https://doi.org/10.5244/C.35.18

Akkaynak, D., Britton, W., Twardowski, M., & Dalgleish, F. (2022). *Towards a lidar-integrated underwater imaging system* [Oral presentation]. Ocean Sciences Meeting, Online. https://www.youtube.com/watch?v=Dil_S8JoMu0

Akkaynak, D., & Treibitz, T. (2018). A revised underwater image formation model. 2018 IEEE/CVF Conference on Computer Vision and Pattern Recognition, 6723–6732. https://doi.org/10.1109/CVPR.2018.00703

Akkaynak, D., & Treibitz, T. (2019). Sea-Thru: A method for removing water from underwater images. 2019 IEEE/CVF Conference on Computer Vision and Pattern Recognition (CVPR), 1682–1691. https://doi.org/10.1109/CVPR.2019.00178

Akkaynak, D., Treibitz, T., Shlesinger, T., Loya, Y., Tamir, R., & Iluz, D. (2017). What is the space of attenuation coefficients in underwater computer vision? 2017 IEEE Conference on Computer Vision and Pattern Recognition (CVPR), 568–577. https://doi.org/10.1109/CVPR.2017.68

Akkaynak, D., Treibitz, T., Xiao, B., Gürkan, U. A., Allen, J. J., Demirci, U., & Hanlon, R. T. (2014). Use of commercial off-the-shelf digital cameras for scientific data acquisition and scene-specific color calibration. *Journal of the Optical Society of America A*, *31*(2), 312–321. https://doi.org/10.1364/JOSAA.31.000312

Allen, J. J., Akkaynak, D., Schnell, A. K., & Hanlon, R. T. (2017). Dramatic fighting by male cuttlefish for a female mate. *The American Naturalist*, *190*(1), 144–151. https://doi.org/10.1086/692009

Arenas, L. M., Troscianko, J., & Stevens, M. (2014). Color contrast and stability as key elements for effective warning signals. *Frontiers in Ecology and Evolution*, *2*, 25. https://doi.org/10.3389/fevo.2014.00025

Bahr, K. D., Severino, S. J. L., Tsang, A. O., Han, J. J., Richards Dona, A., Stender, Y. O., Weible, R. M., Graham, A., McGowan, A. E., & Rodgers, K. S. (2020). The Hawaiian Ko'a card: Coral health and bleaching assessment color reference card for Hawaiian corals. *SN Applied Sciences*, *2*(10), 1706. https://doi.org/10.1007/s42452-020-03487-3

Bell, J. F., Maki, J. N., Alwmark, S., Ehlmann, B. L., Fagents, S. A., Grotzinger, J. P., Gupta, S., Hayes, A., Herkenhoff, K. E., Horgan, B. H. N., Johnson, J. R., Kinch, K. B., Lemmon, M. T., Madsen, M. B., Núñez, J. I., Paar, G., Rice, M., Rice, J. W., Schmitz, N., … Yingling, R. (2022). Geological, multispectral, and meteorological imaging results from the Mars 2020 perseverance rover in Jezero crater. *Science Advances*, *8*(47), eabo4856. https://doi.org/10.1126/sciadv.abo4856

Berman, D., Levy, D., Avidan, S., & Treibitz, T. (2021). Underwater single image color restoration using haze-lines and a new quantitative dataset. *IEEE Transactions on Pattern Analysis and Machine Intelligence*, *43*(8), 2822–2837. https://doi.org/10.1109/TPAMI.2020.2977624

Brown, M. S. (2023). Color processing for digital cameras. In *Fundamentals and applications of colour engineering* (pp. 81–98). John Wiley & Sons, Ltd. https://doi.org/10.1002/9781119827214.ch5

Conde, M. V., Timofte, R., Huang, Y., Peng, J., Chen, C., Li, C., Pérez-Pellitero, E., Song, F., Bai, F., Liu, S., Feng, C., Wang, X., Lei, L., Zhu, Y., Li, C., Jiang, Y., Yong, A., Wang, P., Leng, C., … Ju Jung, Y. (2023). Reversed image signal processing and RAW reconstruction: AIM 2022 challenge report. In L. Karlinsky, T. Michaeli, & K. Nishino (Eds.), *Computer vision—ECCV 2022 workshops* (pp. 3–26). Springer Nature Switzerland. https://doi.org/10.1007/978-3-031-25066-8_1

Ferrara, E. F., Bauer, L., Puntin, G., Bautz, F., Celayir, S., Do, M.-S., Eck, F., Heider, M., Wissel, P., Arnold, A., Wilke, T., Reichert, J., & Ziegler, M. (2026). Red, green, blue color indices as proxy for Symbiodiniaceae cell density and chlorophyll content during coral bleaching. *Limnology and Oceanography: Methods*, *24*(6), e70047. https://doi.org/10.1002/lom3.70047

Griffiths, H. J., Anker, P., Linse, K., Maxwell, J., Post, A. L., Stevens, C., Tulaczyk, S., & Smith, J. A. (2021). Breaking all the rules: The first recorded hard substrate sessile benthic community far beneath an antarctic ice shelf. *Frontiers in Marine Science*, *8*, 642040. https://doi.org/10.3389/fmars.2021.642040

International Organization for Standardization. (2026). *Digital imaging—Image storage—Part 4: Digital negative format* (International Standard ISO 12234-4:2026; 1st ed.). https://www.iso.org/standard/86123.html

Johnsen, S. (2016). How to measure color using spectrometers and calibrated photographs. *Journal of Experimental Biology*, *219*(6), 772–778. https://doi.org/10.1242/jeb.124008

Kim, S. J., Lin, H. T., Lu, Z., Süsstrunk, S., Lin, S., & Brown, M. S. (2012). A new in-camera imaging model for color computer vision and its application. *IEEE Transactions on Pattern Analysis and Machine Intelligence*, *34*(12), 2289–2302. https://doi.org/10.1109/TPAMI.2012.58

King, D. H., Wasley, J., Ashcroft, M. B., Ryan-Colton, E., Lucieer, A., Chisholm, L. A., & Robinson, S. A. (2020). Semi-automated analysis of digital photographs for monitoring east antarctic vegetation. *Frontiers in Plant Science*, *11*, 766. https://doi.org/10.3389/fpls.2020.00766

Levy, D., Peleg, A., Pearl, N., Rosenbaum, D., Akkaynak, D., Korman, S., & Treibitz, T. (2023). SeaThru-NeRF: Neural radiance fields in scattering media. 2023 IEEE/CVF Conference on Computer Vision and Pattern Recognition (CVPR), 56–65. https://doi.org/10.1109/CVPR52729.2023.00014

Maalouf, A., DelPreto, J., Lucas, M., Poetto, S., Andreas, J., Torralba, A., Gero, S., Petri, G., Rus, D., & Gruber, D. F. (2026). Cooperation by non-kin during birth underpins sperm whale social complexity. *Science*, *391*(6792), 1355–1360. https://doi.org/10.1126/science.ady9280

Marom, S., Kiflawi, M., Akkaynak, D., & Holzman, R. (2025). Dynamic color change in the grouper *Variola louti* during interspecific interactions and swimming. *Behavioral Ecology*, *36*(2), araf005. https://doi.org/10.1093/beheco/araf005

Mobley, C. D. (1994). *Light and water: Radiative transfer in natural waters*. Academic Press.

Monfort, T., Cheminée, A., Bianchimani, O., Drap, P., Puzenat, A., & Thibaut, T. (2021). The three-dimensional structure of mediterranean shallow rocky reefs: Use of photogrammetry-based descriptors to assess its influence on associated teleost assemblages. *Frontiers in Marine Science*, *8*, 639309. https://doi.org/10.3389/fmars.2021.639309

Mualem, N., Amoyal, R., Freifeld, O., & Akkaynak, D. (2026). Gaussian splashing enables direct volumetric rendering underwater. *Scientific Reports 16*, 29658. https://doi.org/10.1038/s41598-026-56917-4

Pedersen, N. E., Petrovic, V., Runyan, H., Caldwell, L. T., Edwards, C. B., French, B. J., Hanna Holloway, N. L., Lo, E., Lubarsky, C. A., McCarthy, O. S., Sullivan, C. J., Zglizcynski, B. J., Gleason, A. C. R., Kuester, F., & Sandin, S. A. (2025). Measuring the structure, composition, and change of underwater environments with large-area imaging. *Journal of Visualized Experiments*, *218*, 67700. https://doi.org/10.3791/67700

Rovero, F., & Kays, R. (2021). Camera trapping for conservation. In S. A. Wich & A. K. Piel (Eds.), *Conservation technology*. Oxford University Press. https://doi.org/10.1093/oso/9780198850243.003.0005

Schanda, J. (2007). *Colorimetry: Understanding the CIE system*. John Wiley & Sons.

Schramm, K. D., Harvey, E. S., Goetze, J. S., Travers, M. J., Warnock, B., & Saunders, B. J. (2020). A comparison of stereo-BRUV, diver operated and remote stereo-video transects for assessing reef fish assemblages. *Journal of Experimental Marine Biology and Ecology*, *524*, 151273. https://doi.org/10.1016/j.jembe.2019.151273

Shlesinger, T., Akkaynak, D., & Loya, Y. (2021). Who is smashing the reef at night? A nocturnal mystery. *Ecology*, *102*(10), 1–3. https://doi.org/10.1002/ecy.3420

Siebeck, U. E., Marshall, N. J., Klüter, A., & Hoegh-Guldberg, O. (2006). Monitoring coral bleaching using a colour reference card. *Coral Reefs*, *25*(3), 453–460. https://doi.org/10.1007/s00338-006-0123-8

Stevens, M., Párraga, C. A., Cuthill, I. C., Partridge, J. C., & Troscianko, T. S. (2007). Using digital photography to study animal coloration. *Biological Journal of the Linnean Society*, *90*(2), 211–237. https://doi.org/10.1111/j.1095-8312.2007.00725.x

Troscianko, J., & Stevens, M. (2015). Image calibration and analysis toolbox—A free software suite for objectively measuring reflectance, colour and pattern. *Methods in Ecology and Evolution*, *6*(11), 1320–1331. https://doi.org/10.1111/2041-210X.12439

Winters, G., Holzman, R., Blekhman, A., Beer, S., & Loya, Y. (2009). Photographic assessment of coral chlorophyll contents: Implications for ecophysiological studies and coral monitoring. *Journal of Experimental Marine Biology and Ecology*, *380*(1–2), 25–35. https://doi.org/10.1016/j.jembe.2009.09.004

**How to cite this article:** Akkaynak, D., & Brown, M. S. (2026). Underwater imaging without colour distortions requires RAW capture. *Methods in Ecology and Evolution*, *00*, 1–9. https://doi.org/10.1111/2041-210x.70417